\documentclass[12pt]{article}
\usepackage{graphicx}
\title {Time variability of high energy cosmic rays}
\author {A.D.Erlykin $^{1,1a}$ and A. W. Wolfendale $^{1}$\\
$(1)$ Department of Physics, Durham University, Durham, UK\\
$(1a)$ Permanent address: P N Lebedev Institute, Moscow, Russia}

\date{\today}

\begin{document}
\maketitle

\begin{abstract}
Our model involving cosmic ray acceleration in supernova remnants has been used to 
predict cosmic ray intensities over long periods of time on a statistical basis.  If,
as is highly probable, extensive air showers caused by PeV cosmic rays are needed to 
initiate terrestrial lightning then past dramatic changes in PeV intensities may have 
had important biological effects. 

The model has been used to estimate the manner in which the PeV cosmic ray intensity at
 Earth has varied over the past tens of thousands of years.
\end{abstract}

\section{Introduction}
Over the 4.5By since the formation of the Earth the astronomical environment has been 
variable and, with it, the cosmic ray spectrum.  There are three main sources of 
variability: the Geomagnetic field, the Sun (by way of the solar wind) and the 
presence of nearby cosmic ray (CR) sources.  The first two relate to variations of the 
low energy particles, principally below 10 GeV, and the last-mentioned to all energies.

In addition to an interest in its own right, the variation of CR energy spectrum with 
time has relevance to atmospheric 
properties and thus (perhaps) to the human condition.  Concerning solar effects, the 
most prominent variation is the 11-year 'Solar Cycle'. An effect on terrestrial climate
 is debatable; whereas some (eg Svensmark, 2007) attribute the correlation of CR 
intensity, by way of neutron monitor measurements, with low cloud cover to `cause and 
effect' others, including ourselves (eg Erlykin et al, 2009) differ.  Nevertheless, in 
the upper atmosphere there are genuine signals of changes (ionsphere, ozone) which can 
be attributed to the occasional high fluxes of solar protons.

Our work reported here relates to higher energies than those concerned with solar 
effects, specifically $10^{14}$eV and above.  There is possible relevance to the 
atmosphere (and humans) by way of the likely role of such particles in the 
initiation of lightning (eg Gurevich and Zybin, 2001, Chubenko et al., 2009, Gurevich 
et al., 2009, Chilingarian et al., 2009). The idea is that the leader lightning stroke 
is initiated by runaway electrons which are part of extensive air showers (EAS). The 
references quoted include observed coincidences between EAS and lightning and not just
 the undoubted effect of thunderstorm electric fields on the energies of CR particles 
(which are, themselves, not members of EAS). 

Our own estimates confirm that low energy CR do not correlate with lightning frequency.
The frequency of lightning strokes vs. geomagnetic rigidity cut-off has the best fit 
slope index of 0.23$\pm$0.14 to be compared with -0.8 expected if the neutron monitor
counting rate was relevant.

On the other hand our estimates of the zenith anglular distribution of lightnings is 
close to the expected for EAS. The mean zenith angle of the leader stroke is 17$\pm$3
degrees to be compared with 19 degrees for EAS. The mean zenith angle for the main 
strokes is 20$\pm$2 degrees.  

The whole question of the electrical 
conditions of the atmosphere including its most dramatic manifestation (lightning) is 
tied up with CR insofar as they represent an important source of ions near ground level
 and the major source at altitudes above a few km. Tinsley et al., 2007 and others 
have pointed out the great importance of the 'global electric circuit' - to which CR 
contribute considerably - even when the changes considered have been small. Changes 
consequent upon 'our' very large changes in CR intensity could be profound.

Lightning has, conceivably, played a role in the evolution of life.  Starting with 
pre-life, the work of Miller and Urey 
(Miller, 1953) involving the passage of electrical discharges through a 'pre-biotic 
soup' of appropriate chemicals caused 
quite complex molecules to be generated (monomers, RNA etc) which were necessary 
pre-cursors of elementary life.  Thus, 
lightning could, conceivably, have provided the required discharges.

Later, when `life' was advanced, lightning could have had an effect on evolution by 
virtue of the obnoxious NO$_x$ 
(NO and NO$_2$) produced.  Even now, some 20\% of NO$_x$ comes from lightning - 
much higher lightning rates could have been important.

Even if none of the above effects turn out to be important, a knowledge of the past 
history of the intensity of high energy CR (HECR), by which we mean 10$^{14}$eV and 
above, is of considerable interest.

We start with an analysis of the time variation on a statistical basis using results 
provided by us earlier (Erlykin and 
Wolfendale, 2001a), and based on our supernova remnant model of CR acceleration 
(Erlykin and Wolfendale, 2001b).  We then 
go on to make an examination of the manner in which the HECR intensity on earth has 
varied in the recent past - some 30,000 y - 
assuming that our Single Source Model of the `knee' in the spectrum at 
$\sim 3\cdot 10^{15}$ eV (Erlykin and Wolfendale, 1997) is correct.

\section{Variations of cosmic rays over the past million years using the results of a 
Monte Carlo analysis}
\subsection{Time Profiles as a function of energy}
Figure 1 shows a typical time-profile for different energies.  Protons are assumed in 
the calculations but the results can 
be applied to other nuclei by simple rigidity-transformation.  
\begin{figure}[htb]
\begin{center}
\includegraphics[height=9cm,width=14cm]{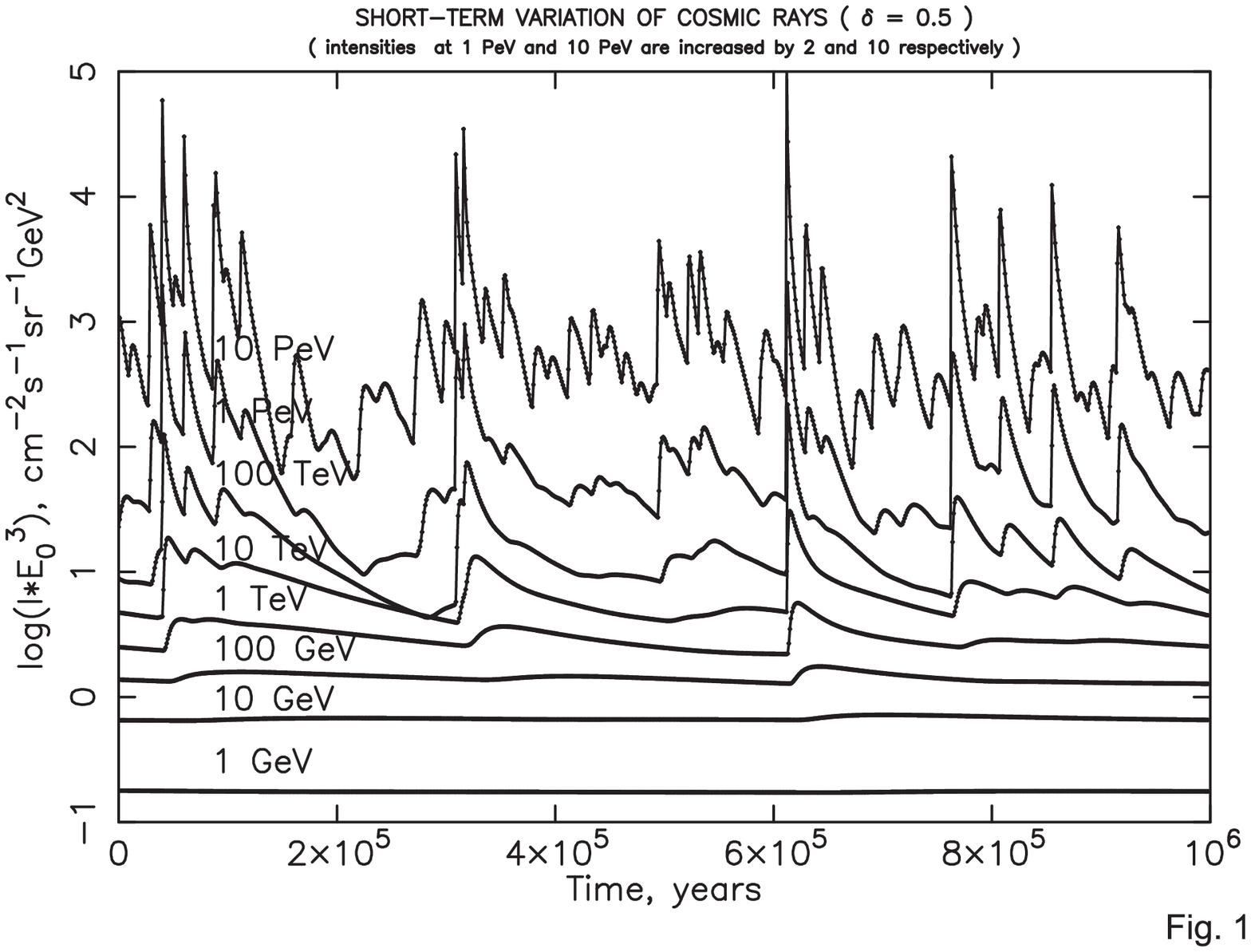}
\caption{\footnotesize Short-term variations of cosmic rays over a period of 1 million 
years, using our statistical model (Erlykin and Wolfendale, 2001a).  The time interval 
is 1 ky.  The results relate to the EW model with an energy dependent diffusion 
coefficient having exponent $\delta = 0.5$ and supernova remnants 
accelerating CR protons up to a maximum energy of 10 PeV.
The intensities at 1 PeV and 10 PeV are displaced upwards by 2 and 10 respectively for 
ease of discrimination.}
\label{fig:fig1}
\end{center}
\end{figure}
It will be noted that in addition to the rare upward excursions, which are particularly
 marked at the highest energies (taken here as 10 PeV) there are long periods - by 
chance - when the average level is well below the mean. Larger variations of CR at 
higher energies are the consequence of the stochastic nature of supernova explosions in
space and time. Recent and closer explosions produce CR with flatter spectrum not 
distorted yet by propagation effects ( the slope index is $\sim$2.1 compared with 
$\sim$2.7 for the background ) and their excess contribution is more pronounced at 
higher energies.

Starting with the positive excursions, we define `peaks' above nearby minima and give 
the usual log N - log S plot where S is the intensity of a peak and N is the number of 
times such a peak intensity, or bigger, is achieved.  The result is shown in Figure 2.
\begin{figure}[htb]
\begin{center}
\includegraphics[height=9cm,width=14cm]{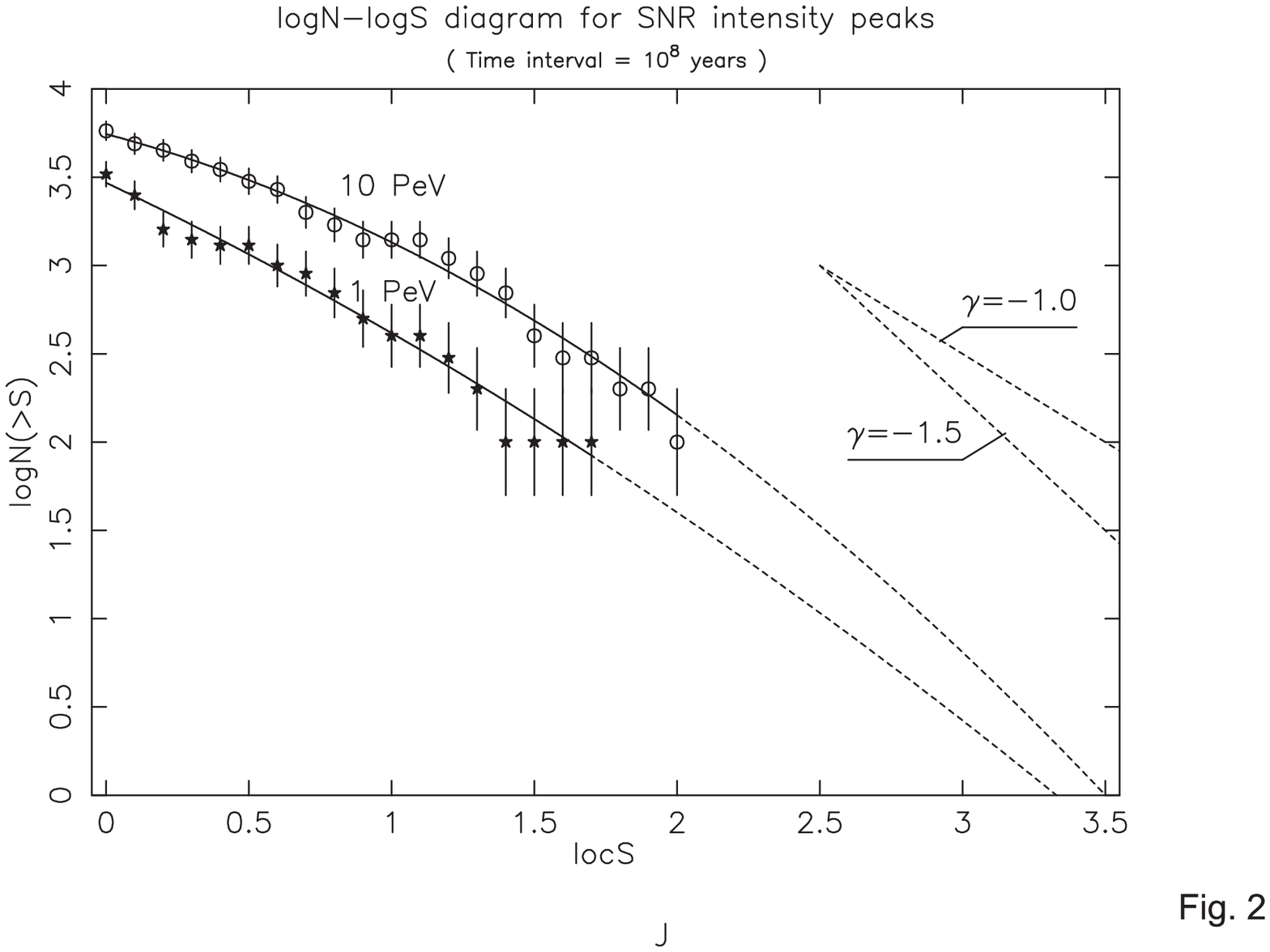}
\caption{\footnotesize The `log N - log S' plot for peak heights from Figure 1, 68 in 
all.  Each peak has height `S' = the ordinate in Figure 1 minus the previous minimum.  
N($>$S) is the number of peaks per My of height $>$S.
The lines are simple parabolic fits to the points.  Importantly, in the middle region 
(log S $\sim 1$) they have slope $\gamma = 1$, appropriate to a 2-dimensional 
distribution of sources, whereas at high values of S they are as appropriate for a 3-D 
distribution of sources.}
\label{fig:fig2}
\end{center}
\end{figure}
We remember that the data are binned in 1000 y.  The lines drawn in Figure 2 are simple
 parabolic `best-fits'.  In the `middle region', say log S = 1 or 2, the shape should 
follow a line of slope 1 since, here, we are dealing with, essentially, a 
two-dimensional distribution of sources (SNR), these being further away than the 
half-thickness of the SNR distribution (hwhm $\simeq 250$ pc).  At S values below 1 
the curvature arises from loss of small S-values due to `source-confusion'.  
Eventually, above log S = 2, the slope should tend to -1.5 because some of the sources 
will be nearer than 250 pc and the distribution of relevant sources tends to 
three-dimensional.

Of particular interest is the extension to cover a period of $10^{8}$ years, the likely
 period $\simeq 3.8 $- $3.9$ Gy ago (or somewhat longer) when pre-biotic life formed on
 Earth.  We note that of order one peak would occur for energy 10 PeV and above with 
intensity some 3,000 times the datum. Taking the median value of log I - 1.69 the 
enhancement is a factor of about 60.  Such an enhanced intensity would continue for a 
few thousand years.

The model ignores the fact that SN II (the type responsible for the CR in question)
are formed predominantly in Galactic Spiral Arms, which are crossed by the solar system 
every 10$^8$ years. The result is simply to re-distribute the pattern in time, without 
having an effect on the mean frequency of the large intensity excursions. 
\subsection{Temporal effects}
It is of relevance to examine the fraction of time for which the CR intensity would be 
above and below certain limits over our `standard' period of $10^{6}$ y.  This is given
 in Figure 3 for 10 PeV. 
\begin{figure}[htb]
\begin{center}
\includegraphics[height=9cm,width=14cm]{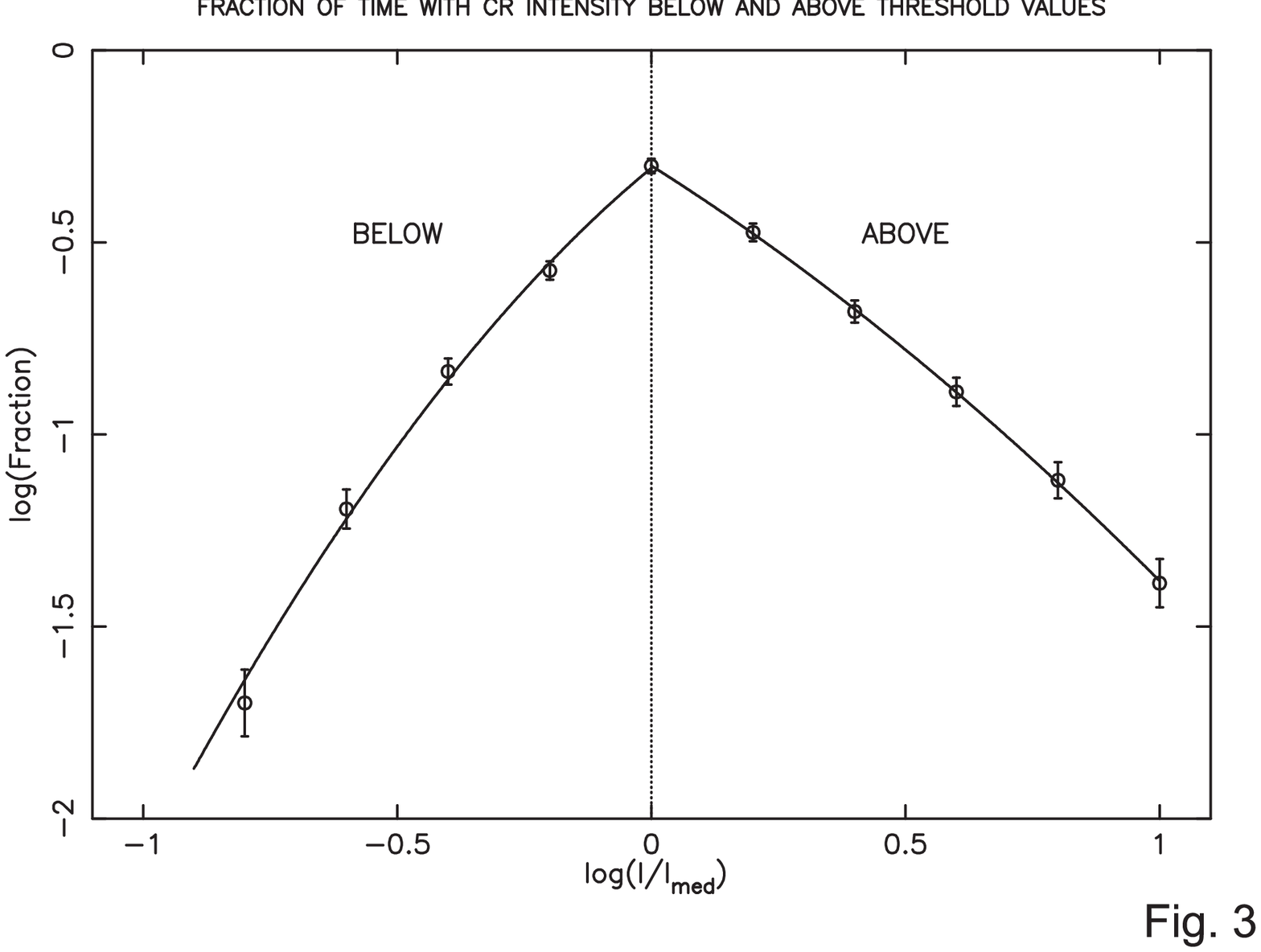}
\caption{\footnotesize Fraction of time (in the My sample) for which the intensity is 
$\tau$ times larger (in logarithmic units) than the overall median value and the 
fraction for which the intensity is $\tau$ times smaller.}
\label{fig:fig3}
\end{center}
\end{figure}
 It will be noted that for 10\% of the time the intensity will be about ten times the 
median and for 10\% of the time, the intensity would be about one quarter of the median.

\subsection{Short-term variations for 10,000 year bins}
Figure 4 shows the equivalent to Figure 1 for time bins which are ten times that used 
previously, viz 10,000 years.  The 10 PeV peaks are typically 5 times smaller than in 
Figure 1.  The equivalent of Figure 2 would give an enhancement by about a factor 10, 
lasting 10,000 y every 100 My.
\begin{figure}[htb]
\begin{center}
\includegraphics[height=9cm,width=14cm]{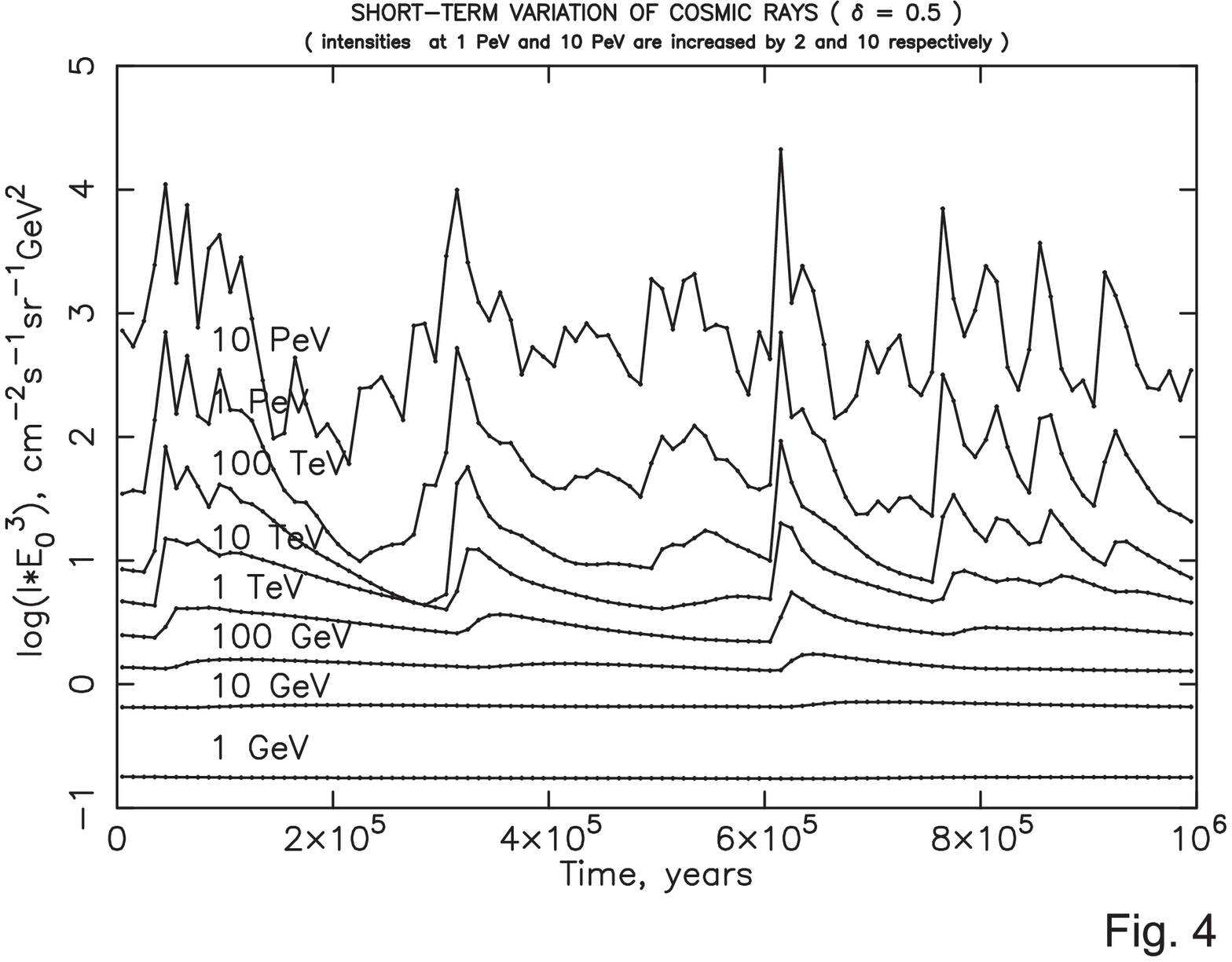}
\caption{\footnotesize As for Figure 1 but for a time interval of 10,000 y.}
\label{fig:fig4}
\end{center}
\end{figure}

\section{The likely intensity in the immediate past}
In our earlier work (Erlykin and Wolfendale, 2003), we identified the `single source', 
responsible for the `knee' in the cosmic ray spectrum as probably being in the distance
 range 250 - 400 pc and being of age in the range 85 to 115 ky.  Figure 5 shows the 
results of our calculations for a distance of 300 pc and the range of ages just 
indicated.  It will be noted that the ratio of the intensity (for 10 PeV) at the peak 
to that at present covers a wide range: from 10 to 1000.  Certainly, in the `recent 
past' (some thousands to tens of thousands of years), the intensity should have been 
significantly higher than at present. It is conceivable that a study of ancient Chinese and Korean records would give useful records of past lightning frequencies.
\begin{figure}[hptb]
\begin{center}
\includegraphics[height=14cm,width=14cm]{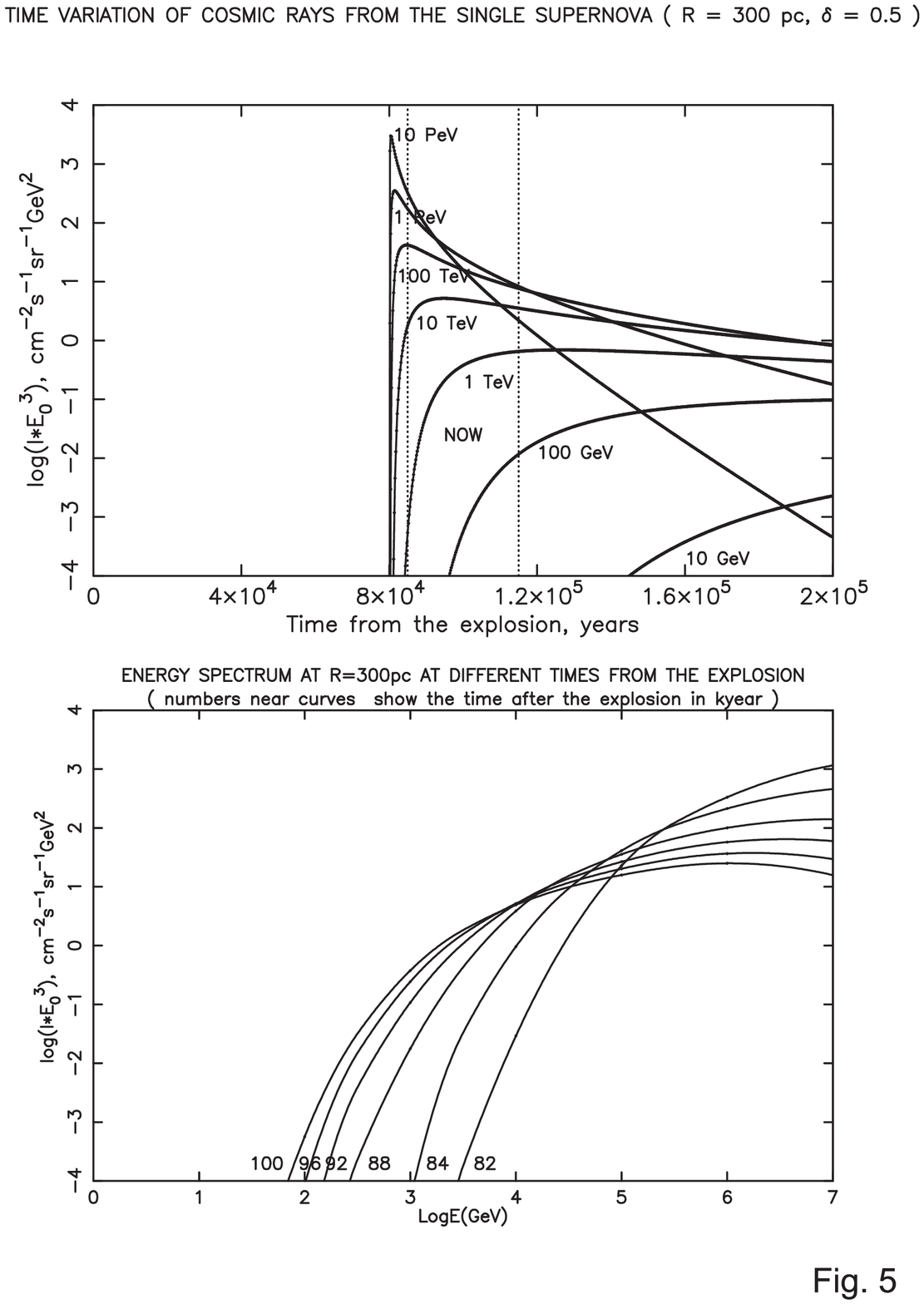}
\caption{\footnotesize The CR intensity from a single SN at 300 pc. Our estimated range
 of the `present time' is indicated by vertical dotted lines in the upper panel.}
\label{fig:fig5}
\end{center}
\end{figure}

\section{Discussion and Conclusions}
The results are, from the CR point of view, straightforward: considerable fluctuations 
in PeV CR intensities should occur over long periods of time (My). The corresponding 
changes in the total CR flux would be very small. At high energies, in fact, the 
variations will be bigger than quoted if, as seems possible, the diffusion coefficient 
in the `local bubble' in the interstellar medium, in which we reside, is higher than 
the conventional one - for a uniform interstellar medium - adopted in our calculations.

It can be remarked that, since most of the fluctuations are stochastic and geometrical 
in origin, CR production by other types of `discrete' sources, such as pulsars, would 
give rather similar results.The very close SNR responsible for the dramatic upward 
UHECR intensity fluctuations, would also have dramatic 'gamma ray flashes' which could 
also have a dramatic effect on the Earth, not least on the ozone layer (eg. Wdowczyk and
Wolfendale, 1977 and references therein).

Turning to the relevance of the results to lightning, and to possible biological 
effects in the 100 My window for life creation, the considerable increase in 10 PeV 
intensity for some tens of thousand years, with the presumed increased lightning rates - could have played a part in the pre-biotic life generation.

At later stages, when life was evolving, the occasional lightning excesses with 
increased production of NO$_x$ could have had pronounced positive effects on 
vegetation and negative effects on humans.  However, evolutionary spurts may have 
occurred for these long periods when the 10 PeV intensity was low.

The claimed PeV increase in the past could (over,say, 5000 years) conceivably be found 
in historical records of changes in lightning rates.

We know of no work related to global circuit caused by dramatic changes in the HECR 
rate. There may be other consequences beside a (presumed) change of the frequency of 
lightning strokes such as their individual strengths - and other atmospheric phenomena.
\section{Acknowledgements}
The Physics Department of Durham University is thanked for the provision of excellent 
facilities. The authors are grateful to the Kohn Foundation for supporting this work.
\section{References}
\begin{itemize}
\item[1] Chilingarian A., Daryan A., Arakelyan K., Reymers A. and Melkumyan L., 
Thunderstorm correlated enhancements of cosmic ray fluxes detected at Mt.Aragats, Proc.
 of Int. Symp. FORGES 2008, Nor-Amberd, Armenia, ed. by A.Chilingarian, (2009), 121
and private communication (2009)
\item[2] Chubenko A P, Karashtin A N, Ryabov V A, Shepetov A L, Antonova V P, Kryukov S
 V, Mitko G G, Naumov A S, Pavljuchenko L V, Ptitsyn M O, Shalamova S Ya, Shlyugaev Yu 
V, Vildanova L I, Zybin K P and Gurevich A V, Energy spectrum of lightning gamma 
emission, Phys. Lett. A, \textbf{373}, (2009), 2953  
\item[3] Erlykin, A D and Wolfendale A W, A Single source of cosmic rays in the range 
$10^{15}$ - $ 10^{16}$ eV, J.Phys.G. \textbf{23} (1997) 979.
\item[4] Erlykin, A D and Wolfendale A W, `Supernova remnants and the origin of the 
cosmic radiation : II spectral variations in space and time', J.Phys.G. \textbf{27}, 
(2001a) 959.
\item[5] Erlykin, A D and Wolfendale, A W, Supernova remnants and the origin of the 
cosmic radiation : I SNR acceleration models and their predictions, J.Phys.G. 
\textbf{27}, (2001b) 941.
\item[6] Erlykin, A D, Gyalai, G, Kudela, K, Sloan T, and Wolfendale A W,`On the 
correlation between cosmic ray intensity and cloud cover', J. Atmos. Sol-Terr. Phys. 
(2009), doi:10.1016/j.jastp.2009.06.012.
\item[7] Gurevich, A V and Zybin K P, Phys.Usp. 44, (2001), 1119.
\item[8] Gurevich A V, Karashtin A N, Ryabov V A, Chubenko A P and Shepetov A L, 
Non-linear phenomena in ionosphere plasma. The influence of cosmic rays and the runaway
 electron breakdown on the thunderstorm discharges, Physics-Uspekhi, \textbf{179}, 
(2009) 779 (in Russian).
\item[9] Miller S L, A production of amino-acids under possible primitve earth 
conditions, Science \textbf{117} (1953), 528.
\item[10] Svensmark H, Cosmoclimatology : a new theory emerges, News Rev. Astron. 
Geophys., \textbf{48}, (2007), 1.18.
\item[11] Tinsley B A, Burns G B and Zhou Li Min, The role of the global electric 
circuit in solar and internal forcing of clouds and climate, Advances in Space 
Research, \textbf{40} (2007) 1126.
\item[12] Wdowczyk J and Wolfendale A W, Cosmic rays and ancient catastrophes, Nature,
\textbf{268} (1977) 510. 
\end{itemize}

\end{document}